\documentclass{PoS}

\usepackage{textcomp} \usepackage{amsfonts} 
\usepackage{amssymb} 
\usepackage{amsmath} 
\usepackage{slashed}
\usepackage{graphicx} 
\usepackage{pifont} 
\usepackage{multirow,lscape}
\usepackage{multicol}
\usepackage{array} 
\usepackage{longtable}
\usepackage[FIGBOTCAP,TABBOTCAP,tight]{subfigure}
\usepackage{verbatim} 
\usepackage{bm} 
\usepackage{comment}
\usepackage{xcolor} 
\usepackage{url} 

\graphicspath{{figures/}} 

\providecommand{\GeV}{\,\mathrm{GeV}}
\providecommand{\MeV}{\,\mathrm{MeV}}
\providecommand{\fm}{\,\mathrm{fm}}

\title{Tuning the hopping parameter in the Oktay-Kronfeld action for
  charm and bottom quarks on a MILC HISQ ensemble}

\ShortTitle{Tuning $\kappa$ in the Oktay-Kronfeld action} 

\author{ Hwancheol Jeong, Weonjong Lee, 
  Jaehoon Leem, \speaker{Sungwoo Park}
  \\ 
  Lattice Gauge Theory Research Center, CTP, and FPRD, \\
  Department of Physics and Astronomy, 
  Seoul National University, Seoul, 151-747, South Korea\\
  E-mail: \email{wlee@snu.ac.kr}} 

\author{Tanmoy Bhattacharya, Rajan Gupta, Yong-Chull Jang\\ 
  T-2, Theoretical Division, \\
  Los Alamos National Laboratory, Los Alamos, NM 87545, USA\\ 
  E-mail: \email{rg@lanl.gov}}

\author{LANL/SWME Collaboration}

\abstract{The first step in the calculation of semi-leptonic form
  factors in the decay of heavy mesons is the tuning of the hopping
  parameter $\kappa$ for the charm and bottom quark masses.  Results
  for the Oktay-Kronfeld (OK) action are presented for one $N_f=2+1+1$
  HISQ ensemble generated by the MILC collaboration at $a\approx
  0.12\fm$ and $M_\pi\approx 310$~MeV.  Estimates of hyperfine
  splitting of heavy-light and heavy-heavy mesons are presented and
  the inconsistency parameter is evaluated.  }
\FullConference{
34th Annual International Symposium on Lattice Field Theory\\ 
24-30 July 2016\\ 
University of Southampton, UK} 

\begin{document} 

\section{Introduction} 
There are two independent methods to extract $V_{cb}$ with $B$-meson
decays.
One is the heavy quark expansion method based on QCD sum rules with
inclusive $B$-meson decays: $\bar{B}\to X_{c}\ell\bar{\nu}$,
and the other is the lattice QCD method to calculate the semileptonic
form factors in the analysis of the exclusive $B$-meson decays:
$\bar{B} \to D^{(\ast)} \ell \bar{\nu}$.
There exists about $3\sigma$ tension in $|V_{cb}|$ between the
inclusive and exclusive decay channels \cite{ Bailey:2015tba,
  Bailey:2016dzk}.
The future experiment at KEK (Belle 2) will increase statistics
for $B$-meson decays dramatically (by a factor of 50).
It is time to improve the lattice results of semileptonic 
form factors for the exclusive $B$-meson decays.
Since the dominant error in lattice QCD results for $|V_{cb}|$ comes
from the heavy quark discretization, we simulate the Oktay-Kronfeld
(OK) action~\cite{ Oktay:2008ex}, a highly improved version of the
Fermilab formulation.

If we use the OK action instead of the clover action (the original
action of the Fermilab formulation \cite{ElKhadra:1996mp}), then the
power counting estimate suggests that the discretization error due to
charm quarks can be reduced from 1.0\% (clover) down to 0.2\% (OK)
for the semileptonic form factor for the $\bar{B} \to D^* \ell
\bar{\nu}$ decay at zero recoil.
The OK action is improved to $\mathcal{O}(\lambda^3)$ in HQET power
counting, and $\mathcal{O}(v^6)$ in NRQCD power counting, while the
clover action is improved to $\mathcal{O}(\lambda^2)$ in HQET and to
$\mathcal{O}(v^4)$ in NRQCD.
One drawback is that the OK action takes significantly more
computing resources (by a factor of $\approx 50$) to calculate its
propagator.
We measured heavy-light (HL) and heavy-heavy (HH) meson spectra to probe the
improvement by the OK action, and the inconsistency parameter and
hyperfine splitting showed clear improvement \cite{Bailey:2016kbw}.

In this paper, we tune hopping parameters using the physical $B_s$ and
$D_s$ meson spectrum on the coarse MILC HISQ ensemble at $a \approx
0.12 \,\mathrm{fm}$.

\section{Simulation Details} 
We use the coarse ($a\approx 0.12$\,fm) ensemble of the MILC HISQ lattices
\cite{Bazavov:2012xda}.
The lattice geometry is $24^3\times 64$.
The tadpole improvement coefficient is $u_0=0.86372$ from the
plaquette Wilson loop.
The sea quark masses are $am_\ell=0.0102$ for light quarks,
$am_s=0.0509$ for the strange quark, and $am_c=0.635$ for the charm
quark.  For the HL mesons, $B_s^{(\ast)}$ and $D_s^{(\ast)}$,
we use the HISQ action for the strange quark, and the OK action for
charm and bottom quarks.
Heavy quark propagators are generated using an optimized BiCGStab
inverter \cite{Jang:2013yqa}.
In the OK action, the tadpole improved bare quark mass $m_0$ is
related to the hopping parameter $\kappa$ as follows,
\begin{align}
  am_0 = \frac{1}{2u_0\kappa} -(1+3\zeta r_s + 18c_4)
\end{align}
where $c_4$ is the tree-level matching coefficient of a dimension-7
operator in the OK action \cite{Oktay:2008ex}.
Here, we set $\zeta=1$ for isotropic lattices and $r_s=1$ as is standard
for the Wilson clover action.
To tune the hopping parameter $\kappa$ to the physical values, we
simulate four $\kappa$ values each for the bottom and the charm quarks as shown 
in Table \ref{table:valence}. 
The parameters of covariant Gaussian smearing used at both the source and sink of
heavy quark propagators to reduce the excited state contamination
\cite{Yoon:2016dij} are given in Table \ref{table:valence}. 
For HISQ valence quarks, we use point source and sink.

\begin{table}[t!]
  \vspace{-7mm}
  \renewcommand{\subfigcapskip}{0.55em}
  \subtable[HISQ parameters]{
  \begin{tabular}{c@{\hskip 1pc}c}
    \hline\hline
    $am_s^v$& $\epsilon$  \\
    \hline
    $0.0509$ & -0.0017468 \\
    \hline\hline
  \end{tabular}    \label{table:valence_HISQ}
  }
  \hfill
  \subtable[Tadpole improved OK parameters]{
  \begin{tabular}{c@{\hskip 1pc}c@{\hskip 1pc}c@{\hskip 1pc}c}
    \hline\hline
    $\kappa$ (bottom) & $\kappa$ (charm) & $\sigma$ & $N_{GS}$\\
    \hline
    0.042, 0.041, 0.040, 0.039 & 0.049, 0.0487, 0.0483, 0.048 & 5 & 60\\
    \hline\hline
  \end{tabular}    \label{table:valence_OK}
  }
  \caption{Parameters used for generating the valence quark propagators.    \protect\subref{table:valence_HISQ}
    $m^v_s$ is set to the physical strange quark mass, and $\epsilon$
    is the coefficient of the Naik term in the HISQ action~\protect\cite{Bazavov:2012xda}.
    \protect\subref{table:valence_OK} $\kappa$ values for the bottom and charm
    quarks. The covariant Gaussian smearing parameters $\sigma$ 
    and $N_{GS}$ are defined in Ref.~\cite{Yoon:2016dij}.
  }
  \label{table:valence}
\end{table}

We calculate HL and HH meson correlators on 500
configurations using 6 sources for bottom quarks and 3 sources for
charm quarks. 
We use jackknife resampling to estimate the statistical error.
We fix the time separation between sources to $\Delta t = 6$.
We choose the initial source time slice randomly for each
configuration.
We use 11 different momentum projections for the two-point 
meson correlation functions.
To increase the statistics, we use the time reflection symmetry of
the two-point correlation functions.
%

\section{Fits to the Meson Correlators and the Dispersion Relation}

The numerical data for the two-point meson correlators is fit using 
\begin{align}
  f^\text{HL}(t; \mathbf p) &= Ae^{-Et} \big( 1-(-1)^t re^{-\Delta E
    t} \big) + Ae^{-E(T-t)} \big(1-(-1)^t re^{-\Delta
    E(T-t)}\big)\\
  f^\text{HH}(t; \mathbf p) &= Ae^{-Et}+Ae^{-E(T-t)}.
\end{align}
The HL meson correlator, $f^\text{HL}$, has 4 fit
parameters: the ground state energy and amplitude ($E$, $A$), an
amplitude ratio ($r=A^p/A$), and energy difference ($\Delta E=E^p-E$),
where the superscript $p$ stands for the opposite parity partner state 
that is present in staggered fermion correlation functions. 
$f^\text{HH}$ is the function used to fit the HH mesons.
The range $12 \le t \le 19$ is used to fit the HL mesons
and $12 \le t \le 16$ for the HH mesons. The ground state energy $E(\mathbf{p})$ is 
then fit using the following dispersion relation:
\begin{align}
  E(\mathbf p)=M_1 + \frac{\mathbf p^2}{2M_2} - \frac{(\mathbf
    p^2)^2}{8M_4^3} - \frac{a^3W_4}{6}\sum_{i=1}^3 p_i^4 \,,
  \label{eq:dispersion}
\end{align}
to obtain $M_1$ the rest mass, $M_2$ the kinetic mass, $M_4$
the quartic mass, and $W_4$ the Lorentz symmetry breaking term.
In both fits we use the full covariance matrix with trivial priors.

\section{Kappa Tuning}
\label{sec:kappa}

We determine the hopping parameters $\kappa_b$ and $\kappa_c$ such
that the kinetic masses are equal to the physical $B_s$ and $D_s$
masses, respectively.
We tune the kinetic mass $M_2$ rather than the rest mass $M_1$.
The form factors and decay constants which we are interested in
are independent of the rest mass $M_1$ in the Fermilab interpretation
of improved Wilson fermions. 

\begin{figure}[t!]
  \vspace{-10mm}
  \renewcommand{\subfigcapskip}{-0.55em}
  \subfigure{\includegraphics[width=0.47\linewidth]
    {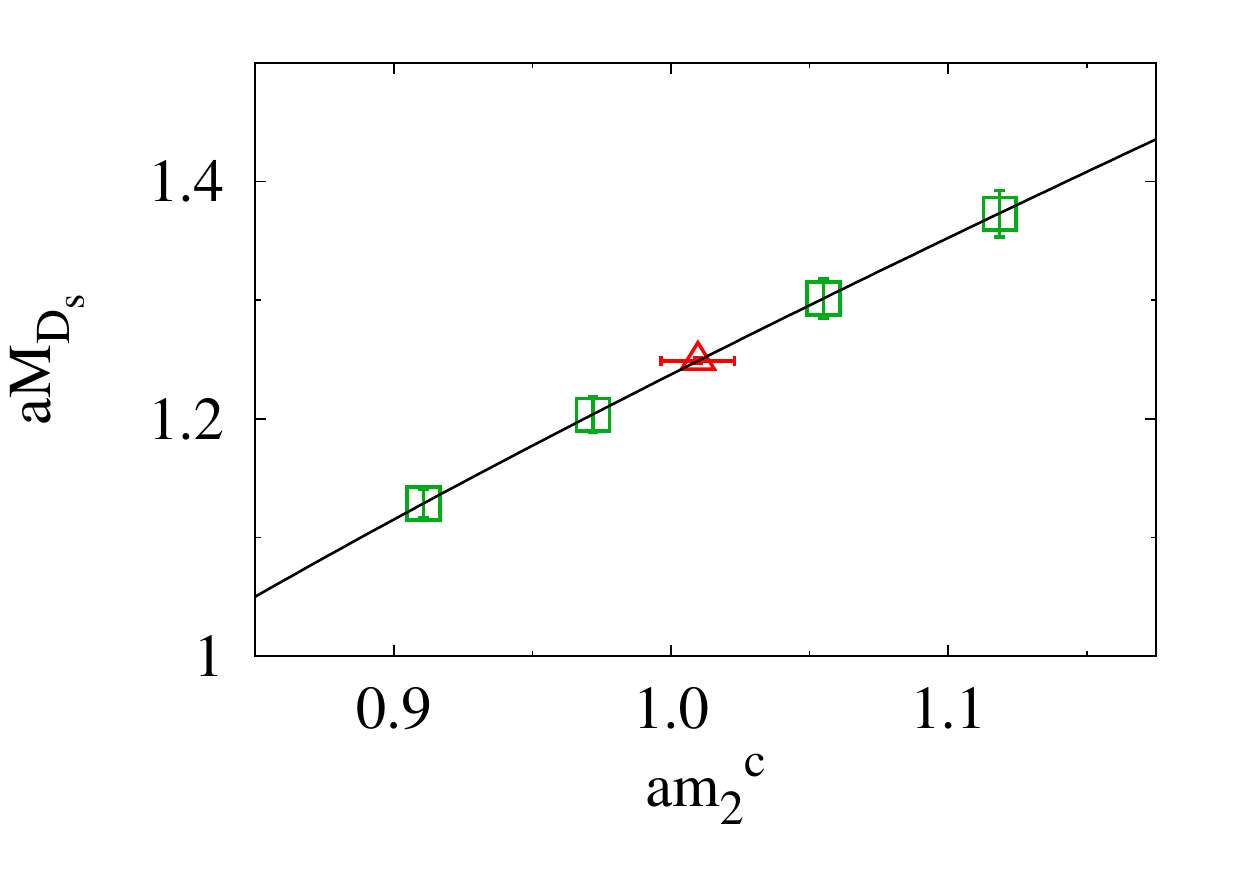}
    \label{fig:kappa-b}
  }    
  \subfigure{\includegraphics[width=0.47\linewidth]
    {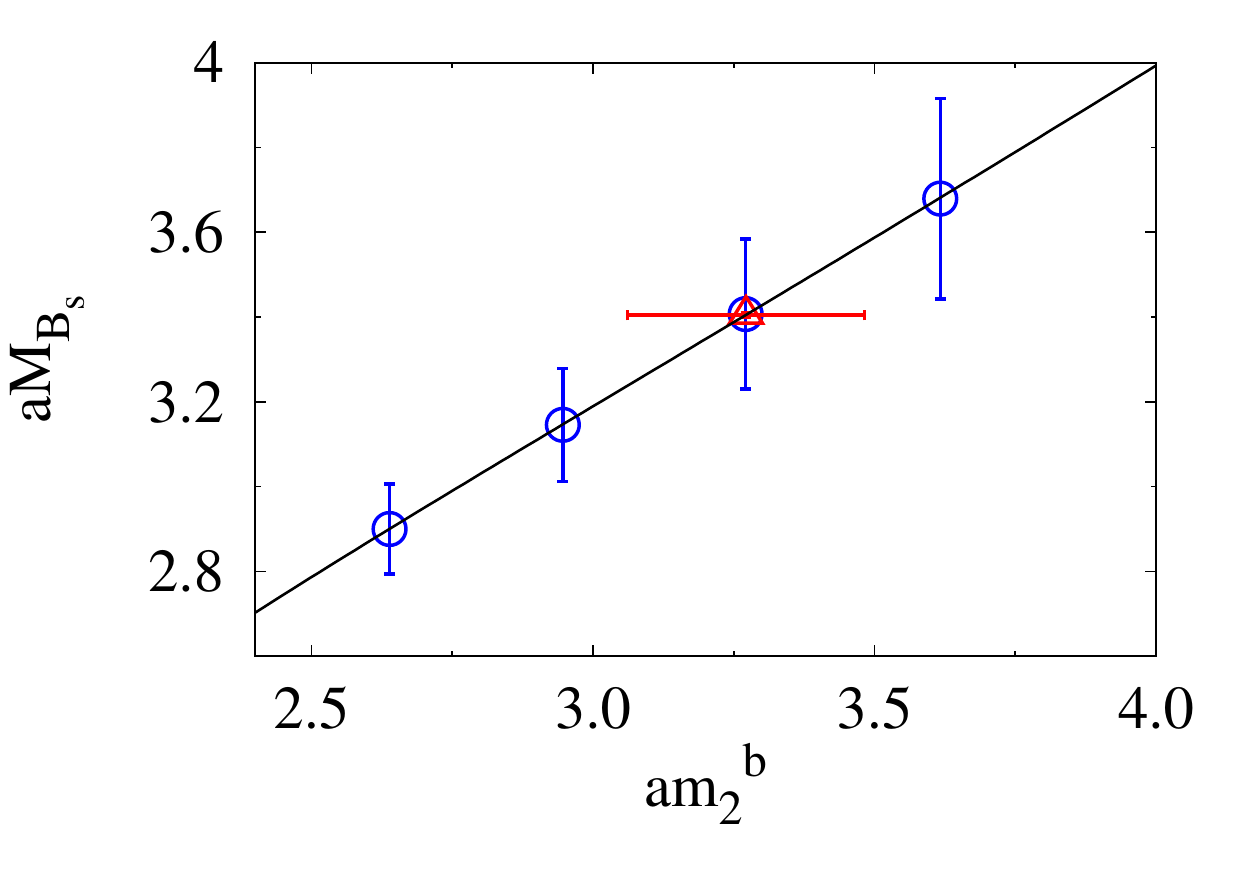}
    \label{fig:kappa-c}
  }    
  \caption{Plot of the pseudoscalar meson mass $D_s$ ($B_s$) versus
    the kinetic quark mass $m_2^c$ ($m_2^b$) defined in
    Eq.~\protect\eqref{eq:def-m2}.  The physical $\kappa_c$ and
    $\kappa_b$ (shown as red triangles and given in Table
    \protect\ref{table:kappa}) are determined by tuning the $D_s$ and $B_s$ pseudoscalar masses 
    to their experimental values.  }
  \label{fig:kappa}
\end{figure}

We use the HQET inspired fitting function for kinetic HL
meson masses,
\begin{align}
  aM_2(\kappa)=d_0+am_2(\kappa)+\frac{d_1}{am_2
    (\kappa)}+\frac{d_2}{(am_2 (\kappa))^2} \label{eq:kappa_fit_function}
  \,,
\end{align}
where $M_2$ is the kinetic mass of the HL meson, and $m_2$ is the
kinetic mass of the heavy quark.
We determine $d_0$, $d_1$ and $d_2$ using the correlated least
$\chi^2$ fitting.
Here, $m_2(\kappa)$ is related to the bare mass
$m_0(\kappa)$ at the tree level as follows,
\begin{align}
  \frac{1}{ am_2} &= \frac{2 \zeta^2}{ am_0 (2+ am_0)} + \frac{ r_s
    \zeta }{1+ am_0 }.
\label{eq:def-m2}
\end{align}
Figure \ref{fig:kappa} shows the interpolation of $m_2$ to the physical
values for the bottom and charm quarks. 
Results for $\kappa_b$ and $\kappa_c$ are summarized in Table
\ref{table:kappa}. 
In Table \ref{table:kappa}, we present the $\kappa$-tuning results using
physical values of the pseudoscalar meson mass ($M_X$), vector meson mass
($M_{X^\ast}$), and the spin-averaged mass $(M_X+3M_{X^\ast})/4$ for $X=B_s$ or $D_s$. 
We find that all the results for $\kappa$ determined from different
spin states are consistent within statistical uncertainty.
We also perform another fit using a simpler fitting function:
$aM_2=d_0+am_2+d_1/(am_2)$, and take the difference in $\kappa$
as the systematic error due to ambiguity in the fitting function.
\begin{table}[!htp]
  \center
  \begin{tabular}{@{\hskip 1pc} l @{\hskip 2pc} l @{\hskip 2pc} l @{\hskip 2pc} l @{\hskip 1pc}}
    \hline\hline
    X & $M_{B_s}^X$ (GeV) & $aM_{B_s}^X$ & $\kappa_b$  
    \\\hline
    pseudoscalar & 5.36682(22) & 3.4051(61) & 0.04000(63)(2)(2) 
    \\
    vector & $5.4154^{+0.0024}_{-0.0021}$ & $3.4360^{+0.0076}_{-0.0074}$ & 0.03932(90)(3)(3)
    \\
    spin-average & $5.4033^{+0.0019}_{-0.0016}$ & $3.4283^{+0.0072}_{-0.0071}$ & 0.03950(80)(3)(4)
    \\ \hline\hline
    X & $M_{D_s}^X$ (GeV) & $aM_{D_s}^X$ & $\kappa_c$ 
    \\ \hline
    pseudoscalar  & 1.96827(10) & 1.2488(23) & 0.048517(63)(9)(1)
    \\
    vector  & 2.1121(4) & 1.3401(26) & 0.048281(163)(12)(2)
    \\
    spin-average  & 2.0761(3) & 1.3173(25) & 0.048346(126)(11)(3)
    \\     
    \hline\hline
  \end{tabular}
  \caption{Results of tuning the $\kappa$ for the bottom ($\kappa_b$)
    and charm ($\kappa_c$) quarks. For converting the experimental $M$
    to $aM$, we use $a=0.12520(22)\fm$~\cite{Bazavov:2014wgs} . In
    $\kappa_{b,c}$, the first error is statistical, the second error
    is propagation of experimental error in $M^X$, and the third error
    is systematic to account for the uncertainty in the fit ansatz.}
  \label{table:kappa}
\end{table}

\section{Inconsistency Parameter}

\begin{figure}[t!]
  \vspace{-5mm}
  \center
  \subfigure{
  \includegraphics[width=0.45\linewidth]{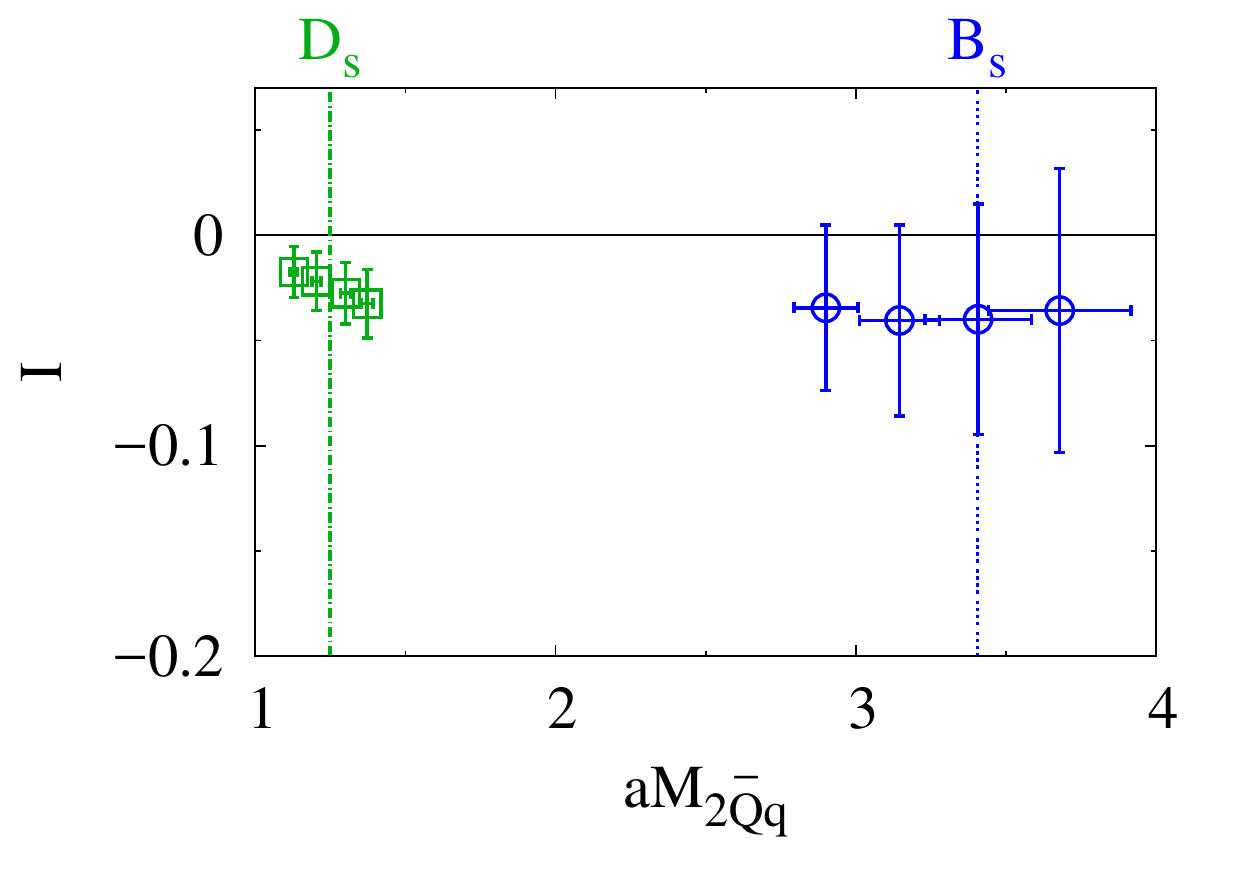}
  }
  \subfigure{
  \includegraphics[width=0.45\linewidth]{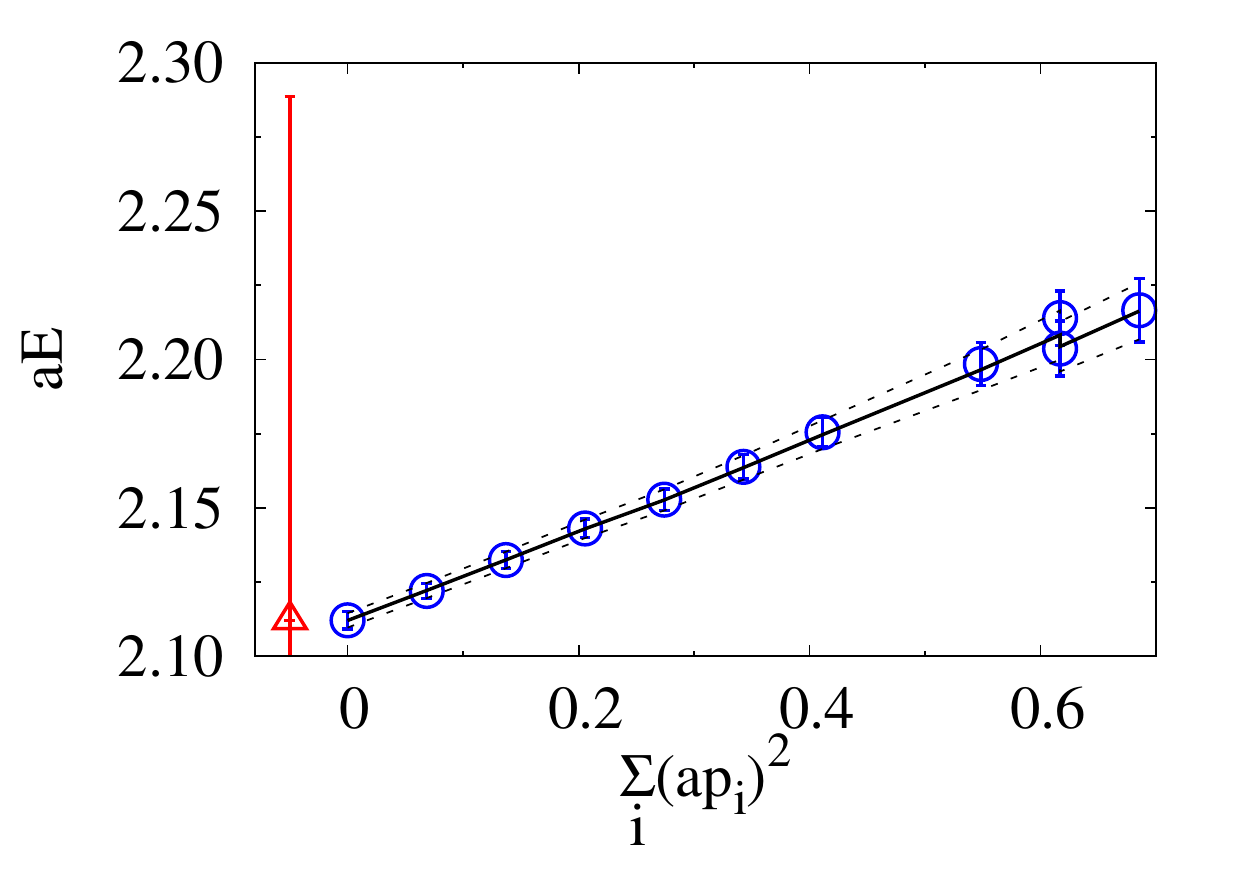}
}
  \caption{(Left) Inconsistency parameter $I$ versus the pseudoscalar
    mass $a M_{2\bar{Q}q}$ for charm (green squares) and bottom (blue
    circles) quarks. (Right) Dispersion relation for the HL bottom
    meson with $\kappa=0.040$. The rest mass, $M_1=2.112(2)$, is given
    by the blue circle at $\bm{p}=0$. The kinetic mass,
    $M_2=3.408(176)$, is extracted from the fit and shown by the red
    triangle translated to $E= 2.112$. Note that the determination
    of the error in $M_2$ is much larger than in $M_1$.    
}
  \label{fig:incon}
\end{figure}

The inconsistency parameter $I$ \cite{Collins:1995yg, Kronfeld:1996uy}
is used to see $O(\mathbf{p}^4)$ improvement in the OK action.
Let us use $Q$ for heavy quarks and $q$ for light quarks, and
define $\delta M \equiv M_{2}-M_{1}$ as the difference between
the kinetic and rest masses.
Then the inconsistency parameter $I$ is
\begin{align}
  I\equiv \frac{2\delta M_{\bar Q q} -(\delta M_{\bar Q Q}+\delta
    M_{\bar q q})}{2M_{2\bar Q q}} 
  = \frac{2\delta B_{\bar Q q} -(\delta B_{\bar Q Q}+\delta
    B_{\bar q q})}{2M_{2\bar Q q}} 
\end{align}
Here the binding energies $B_{1,2}$ are
\begin{align}
  M_{1\bar Q q}= m_{1\bar Q } + m_{1 q} + B_{1\bar Q q }, \qquad
  M_{2\bar Q q}= m_{2\bar Q } + m_{2 q} + B_{2\bar Q q }
\end{align}
for HL mesons.
Here the quark masses $m_{1,2}$ are defined by the quark dispersion
relation, which is similar to Eq.~\eqref{eq:dispersion}.
%
%
We neglect the light quarkonium contribution $\delta M_{\bar q q}$
(and $\delta B_{\bar q q}$).
%
%
%
In Fig.~\ref{fig:incon} we present results for $I$ for pseudoscalar
mesons.
Near the $B_s$ region, $I$ is consistent with the continuum limit, $I=0$,
within the error bars, which indicates a dramatic improvement from that
of the Fermilab action:  $I\approx -0.6$ \cite{Bailey:2016kbw}.

\section{Hyperfine Splittings}

\begin{figure}[t!]
  \vspace{-10mm}
  \renewcommand{\subfigcapskip}{-0.55em}
  \subfigure[Hyperfine splitting for HL mesons]{
    \includegraphics[width=0.49\linewidth]{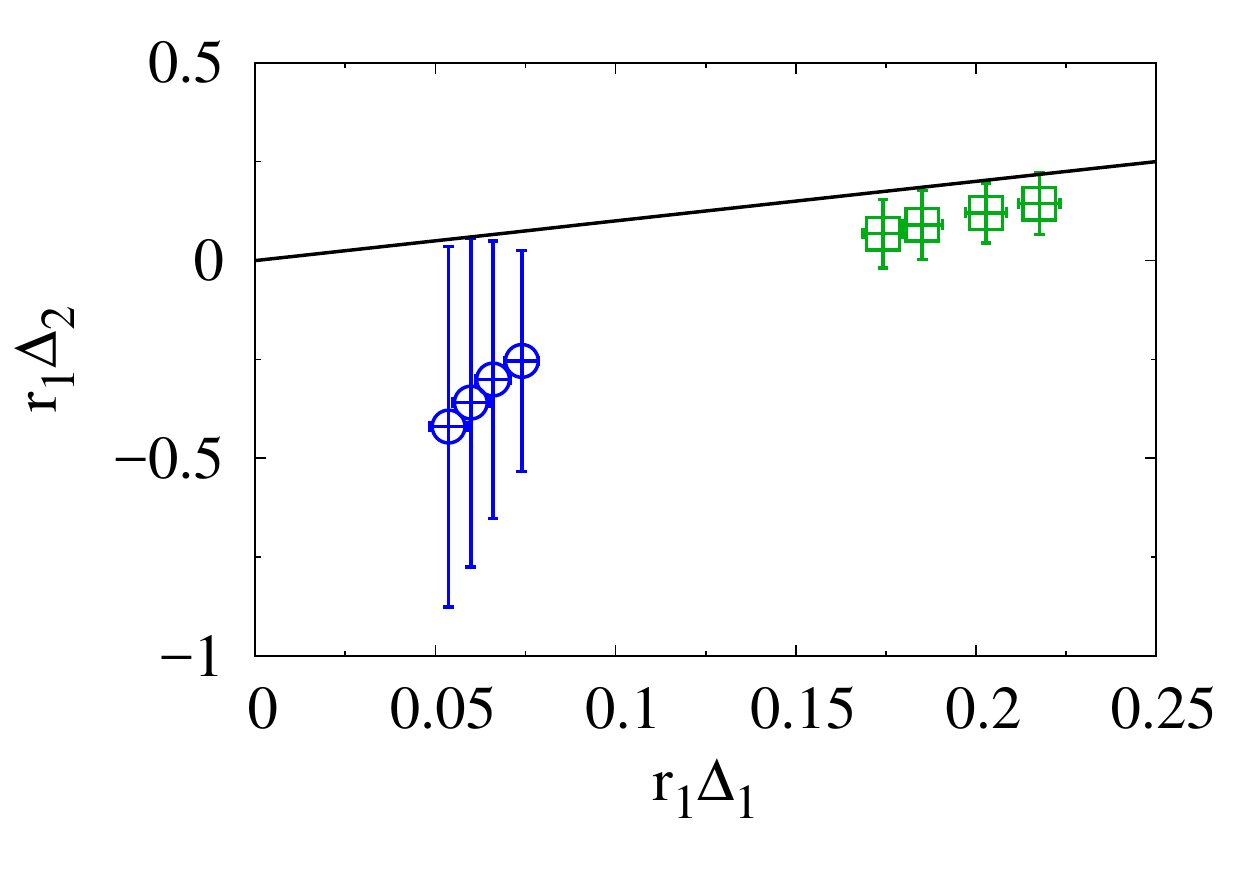}
    \label{fig:hyper-HL}
  }
  \subfigure[Hyperfine splitting for HH mesons]{
    \includegraphics[width=0.49\linewidth]{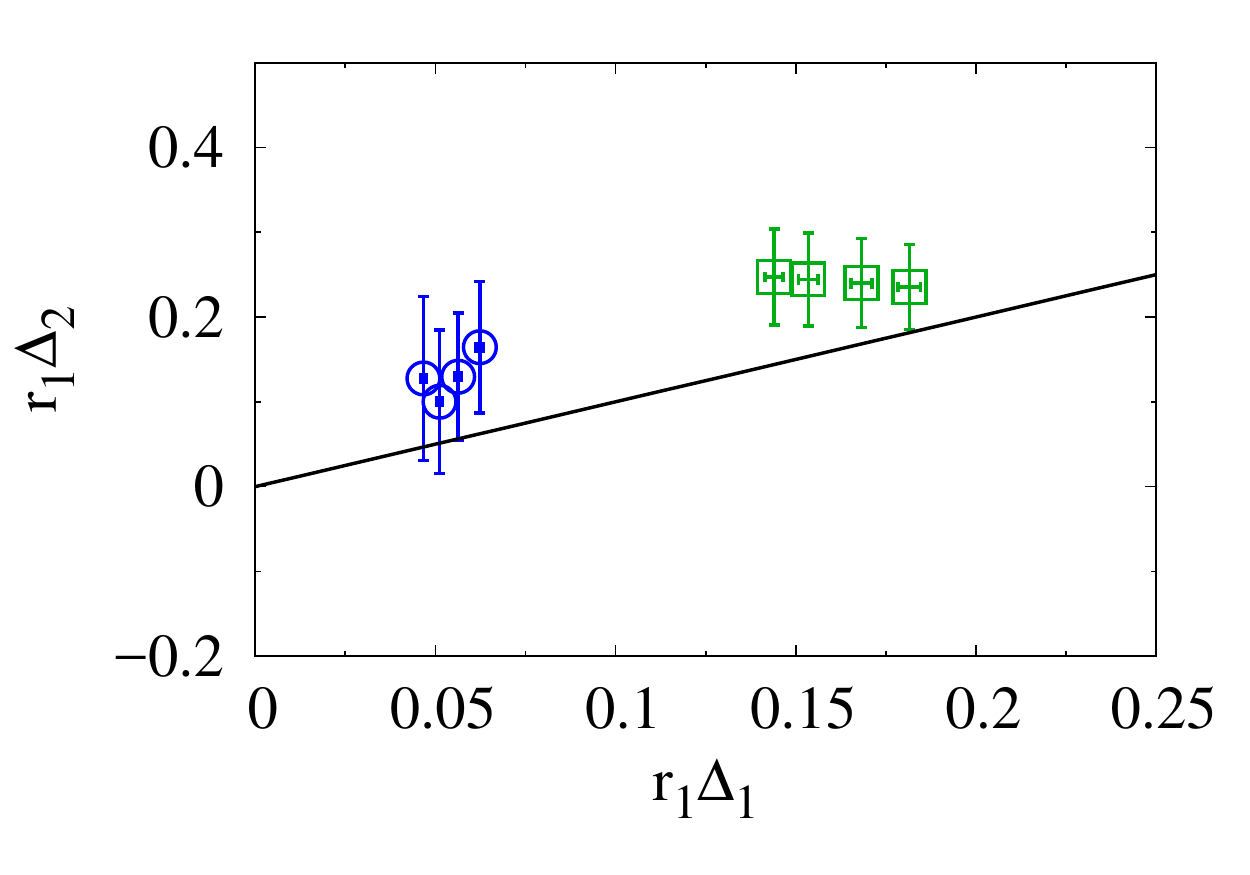}
    \label{fig:hyper-HH}
  }
  \caption{Hyperfine splitting $\Delta_2$ versus $\Delta_1$ for the HL
    (Fig.~\protect\ref{fig:hyper-HL}) and the HH
    (Fig.~\protect\ref{fig:hyper-HH}) mesons in units of $r_1$ taken
    from Ref.~\protect\cite{Bazavov:2012xda}.  The black line represents the
    continuum result, $\Delta_1=\Delta_2$. }
  \label{fig:hyper}
\end{figure}

We define the hyperfine splittings of HL and HH pseudoscalar mesons, $\Delta_1$ and $\Delta_2$ as
\begin{align}
  \Delta_1=M_1^\ast-M_1,\qquad  \Delta_2=M_2^\ast-M_2 \,,
\end{align}
and plot $\Delta_2$ versus $\Delta_1$ in Fig.~\ref{fig:hyper}. As illustrated in 
Fig.~\ref{fig:incon}, $M_2$ has much larger errors than $M_1$ since it is 
extracted from the slope versus momentum. Consequently, $\Delta_2$ has 
larger errors than $\Delta_1$. 

\begin{figure}[t!]
  \vspace{-10mm}
  \renewcommand{\subfigcapskip}{-0.55em}
  \subfigure{
    \includegraphics[width=0.49\linewidth]{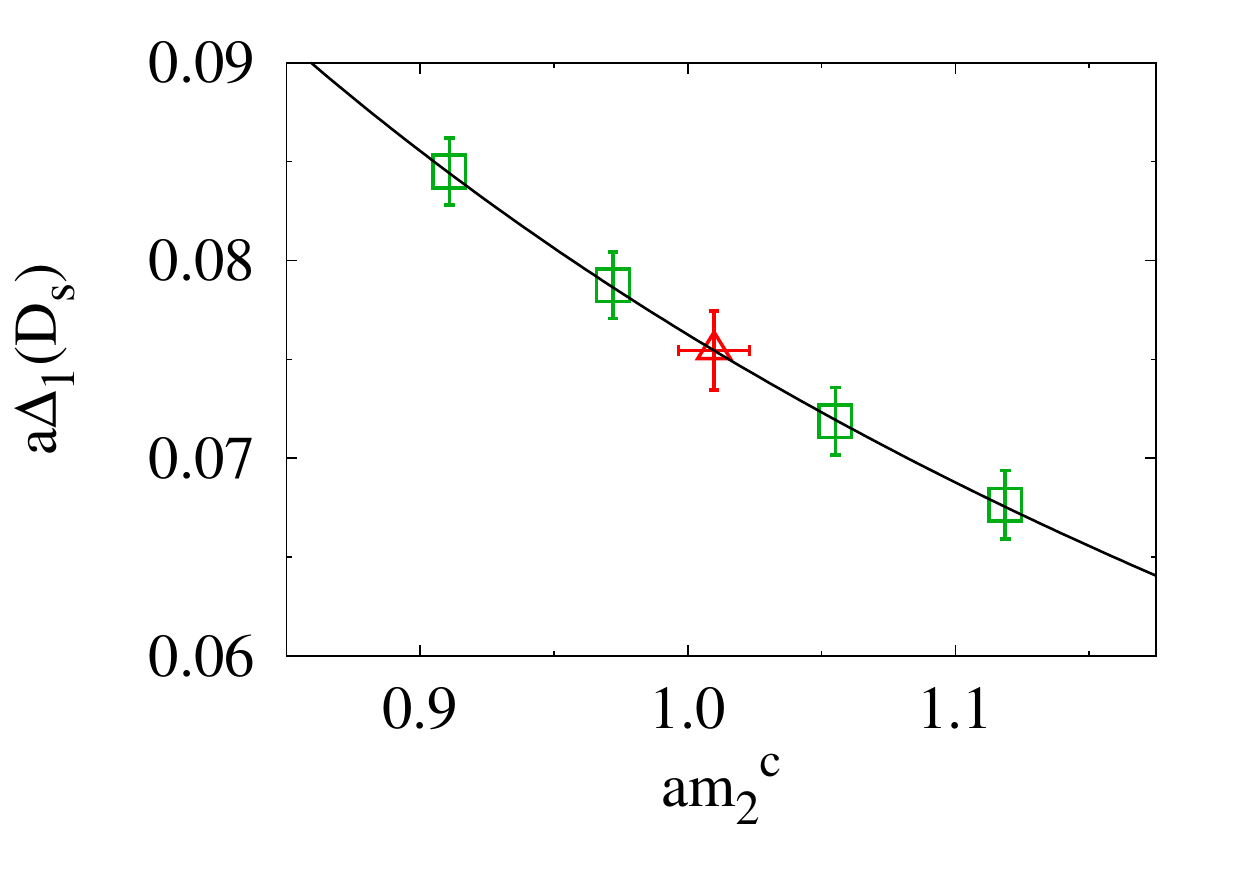}
    \label{fig:hyper_Ds}
  }
  \subfigure{
    \includegraphics[width=0.49\linewidth]{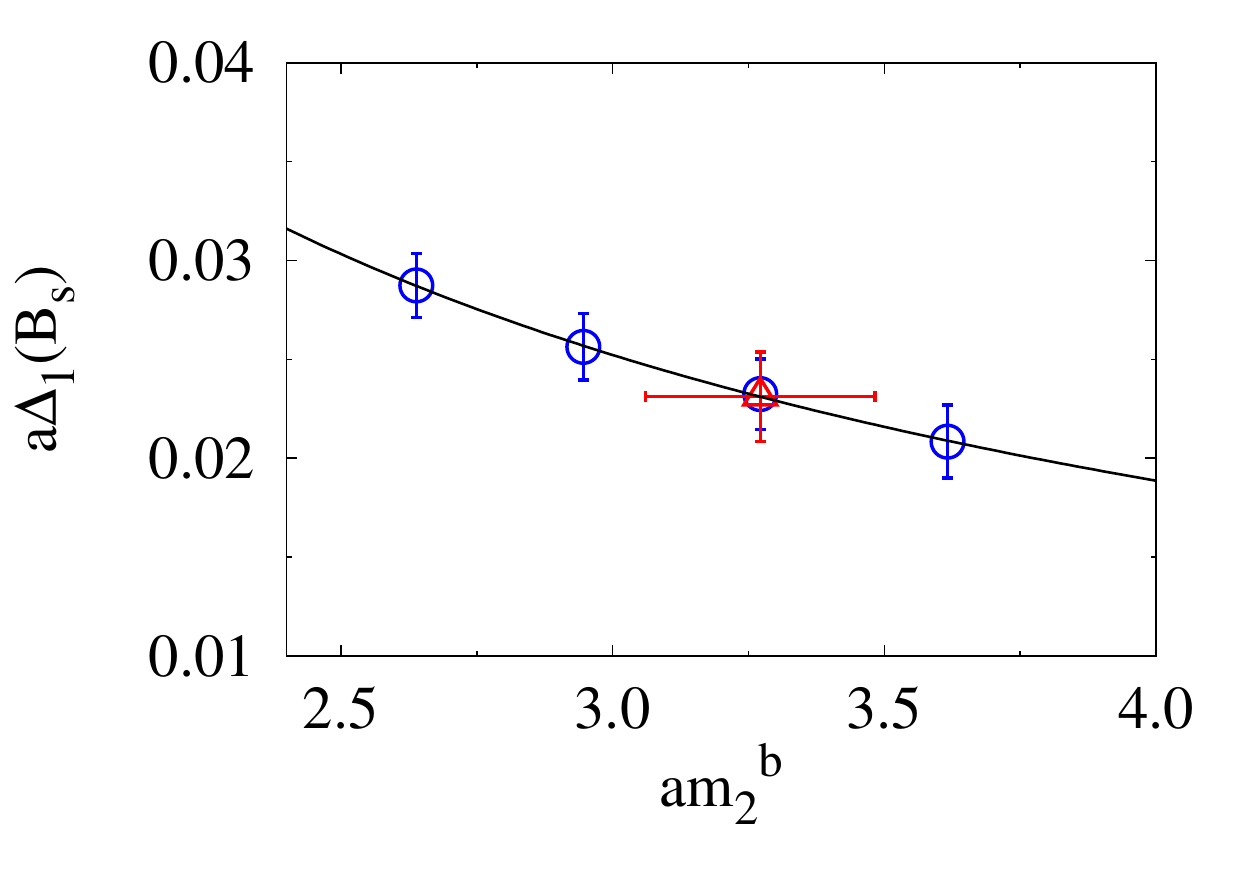}
    \label{fig:hyper_Bs}
  }
  \caption{Plots of $\Delta_1(D_s)$ (left panel) and
    $\Delta_1(B_s)$ (right panel) versus the kinetic masses, $m_2^c$
    and $m_2^b$, of the quarks.  Results at the physical quark masses,
    tuned using the pseudoscalar meson masses, are shown by the red triangles
    and given in Table~\protect\ref{table:hyperfine}.  }
  \label{fig:fit_hyper}
\end{figure}

The HQET expansion for $\Delta_1$ in the
HL meson system is given in Ref.~\cite{Kronfeld:2000ck}:
\begin{align}
  \Delta_1 = M_1^{\ast} - M_1 &= \frac{4\lambda_2}{2m_B} -
  \frac{4\rho_2}{4m_E^2}+ \frac{8
    T_2}{2m_2 2m_B} + \frac{4
    (T_4-T_2)}{4m_B^2} +
  \mathcal{O}\left(\frac{1}{m^3}\right),
\label{eq:Delta1}
\end{align}
where $\lambda_2$, $\rho_2$, $T_2$, $T_4$ are HQET matrix elements
defined in Ref.~\cite{ Kronfeld:2000ck}.
For the OK action, the matching conditions are $m_2=m_B=m_E$~\cite{Oktay:2008ex}. 
Thus, $\Delta_1$ defined in Eq.~\eqref{eq:Delta1} is in terms of the kinetic quark mass, which was used 
to tune the $\kappa$ to the physical value. To analyze $\Delta_1$, we recast 
Eq.~\ref{eq:Delta1} as 
\begin{align}
  a\Delta_{1} = h_0 + \frac{h_1}{am_2} + \frac{h_2}{(am_2)^2}
  + \frac{h_3}{(am_2)^3} \,,
  \label{eq:fit_delta_1}
\end{align}
where $h_1 = 2 a^2 \lambda_2$ and $h_2 = a^3 (-\rho_2 + T_2 +T_4)$. 
Because we have only 4 data points, we set $h_3=0$ in the fits.
Correlated fits, shown in Fig.~\ref{fig:fit_hyper}, give $h_0 = 0$
within statistical uncertainty, consistent with the theoretical
prediction. Our results, with $h_0$ set to zero in the fits are
summarized in Table \ref{table:hyperfine}. The corresponding $h_i$
from fits to $\Delta_2$ were very poorly determined.

We are performing
simulations at other values of the lattice spacing and quark mass in
order to perform the continuum-chiral extrapolation and compare with
the experimental value.

\begin{table}[h!]
  \center
  \begin{tabular}{@{\hskip 0.7pc} l @{\hskip 2pc} l l l l l @{\hskip 2pc} l @{\hskip 0.7pc}}
    \hline\hline
    type & $h_1$ & $h_2$ & $\lambda_2 ~(\GeV^2)$ & $A~(\GeV^3)$ 
    & $\Delta_1~(\MeV)$ & $\Delta_\text{exp}~(\MeV)$ \\\hline
    $B_s$& 0.075(15) & 0.002(35) & 0.093(19) & 0.01(14) & $36.4(3.6)$ & $48.6^{+1.8}_{-1.6}$\\
    $D_s$& 0.0697(45) & 0.0065(33) & 0.0866(59) & 0.025(13) & $118.9(3.4)$ &$143.8\pm 0.4$\\
    \hline\hline
  \end{tabular}
  \caption{Hyperfine splittings, $\lambda_2$ and $A \equiv -\rho_2
    +T_2 +T_4$ for the $D_s$ and $B_s$ mesons at the physical values
    of $\kappa_c$ and $\kappa_b$.  $\Delta_\text{exp}$ is the
    experimental value~\cite{Olive:2016xmw}.}
  \label{table:hyperfine}
\end{table}

\section{Summary and Plan}

We tuned the bottom and charm quark masses using physical values for
$B^{(*)}_s$ and $D^{(*)}_s$ mesons. Estimates from fits to the
pseudoscalar, vector, and spin-averaged mesons masses 
are consistent within their statistical uncertainty (see
Table~\ref{table:kappa}). We used estimates from the pseudoscalar
mesons for the analysis of the hyperfine splittings. These values of
$\kappa_b$ and $\kappa_c$ are now being used to measure the
semileptonic form factors for the exclusive decays $\bar{B}
\rightarrow D^{(*)} \ell \bar{\nu}$.

\acknowledgments
We thank Jon A.~Bailey for helpful comments and suggestions.
The research of W.~Lee is supported by the Creative Research
Initiatives Program (No.~20160004939) of the NRF grant funded by the
Korean government (MEST).
%
%
W.~Lee would like to acknowledge the support from the KISTI
supercomputing center through the strategic support program for the
supercomputing application research (No.~KSC-2014-G2-002).
Computations were carried out in part on the DAVID GPU clusters at
Seoul National University. 
The research of T. Bhattacharya, R. Gupta and Y-C. Jang is
supported by the U.S. Department of Energy, Office of Science of High
Energy Physics under contract number~DE-KA-1401020, the LANL LDRD
program and Institutional Computing.
%

\bibliographystyle{JHEP}
\bibliography{ref} 

\providecommand{\href}[2]{#2}\begingroup\raggedright\begin{thebibliography}{10}

\bibitem{Bailey:2015tba}
{\bf SWME} Collaboration, J.~A. Bailey, Y.-C. Jang, W.~Lee, and S.~Park {\em
  Phys. Rev.} {\bf D92} (2015) 034510,
  [\href{http://xxx.lanl.gov/abs/1503.05388}{{\tt 1503.05388}}].

\bibitem{Bailey:2016dzk}
J.~A. Bailey, W.~Lee, J.~Leem, S.~Park, and Y.-C. Jang {\em PoS} {\bf
  LATTICE2016} (2016) 383, [\href{http://xxx.lanl.gov/abs/1611.00503}{{\tt
  1611.00503}}].

\bibitem{Oktay:2008ex}
M.~B. Oktay and A.~S. Kronfeld {\em Phys. Rev.} {\bf D78} (2008) 014504,
  [\href{http://xxx.lanl.gov/abs/0803.0523}{{\tt 0803.0523}}].

\bibitem{ElKhadra:1996mp}
A.~X. El-Khadra, A.~S. Kronfeld, and P.~B. Mackenzie {\em Phys. Rev.} {\bf D55}
  (1997) 3933--3957, [\href{http://xxx.lanl.gov/abs/hep-lat/9604004}{{\tt
  hep-lat/9604004}}].

\bibitem{Bailey:2016kbw}
{\bf Fermilab Lattice, SWME, MILC} Collaboration, J.~A. Bailey, Y.-C. Jang,
  W.~Lee, C.~DeTar, A.~S. Kronfeld, and M.~B. Oktay {\em PoS} {\bf LATTICE2015}
  (2016) 099, [\href{http://xxx.lanl.gov/abs/1601.04759}{{\tt 1601.04759}}].

\bibitem{Bazavov:2012xda}
{\bf MILC} Collaboration, A.~Bazavov {\em et~al.} {\em Phys. Rev.} {\bf D87}
  (2013), no.~5 054505, [\href{http://xxx.lanl.gov/abs/1212.4768}{{\tt
  1212.4768}}].

\bibitem{Jang:2013yqa}
{\bf Fermilab Lattice, SWME, MILC} Collaboration, Y.-C. Jang, J.~A. Bailey,
  W.~Lee, C.~DeTar, M.~B. Oktay, and A.~S. Kronfeld {\em PoS} {\bf LATTICE2013}
  (2014) 030, [\href{http://xxx.lanl.gov/abs/1311.5029}{{\tt 1311.5029}}].

\bibitem{Yoon:2016dij}
B.~Yoon {\em et~al.} {\em Phys. Rev.} {\bf D93} (2016), no.~11 114506,
  [\href{http://xxx.lanl.gov/abs/1602.07737}{{\tt 1602.07737}}].

\bibitem{Bazavov:2014wgs}
{\bf Fermilab Lattice, MILC} Collaboration, A.~Bazavov {\em et~al.} {\em Phys.
  Rev.} {\bf D90} (2014), no.~7 074509,
  [\href{http://xxx.lanl.gov/abs/1407.3772}{{\tt 1407.3772}}].

\bibitem{Collins:1995yg}
S.~Collins, R.~G. Edwards, U.~M. Heller, and J.~H. Sloan {\em Nucl. Phys. Proc.
  Suppl.} {\bf 47} (1996) 455--458,
  [\href{http://xxx.lanl.gov/abs/hep-lat/9512026}{{\tt hep-lat/9512026}}].

\bibitem{Kronfeld:1996uy}
A.~S. Kronfeld {\em Nucl. Phys. Proc. Suppl.} {\bf 53} (1997) 401--404,
  [\href{http://xxx.lanl.gov/abs/hep-lat/9608139}{{\tt hep-lat/9608139}}].

\bibitem{Kronfeld:2000ck}
A.~S. Kronfeld {\em Phys. Rev.} {\bf D62} (2000) 014505,
  [\href{http://xxx.lanl.gov/abs/hep-lat/0002008}{{\tt hep-lat/0002008}}].

\bibitem{Olive:2016xmw}
K.~A. Olive {\em Chin. Phys.} {\bf C40} (2016), no.~10 100001.

\end{thebibliography}\endgroup


\end{document}